\documentstyle[aps
,epsfig%
,preprint%
]{revtex}
\begin{document}
\draft
\preprint{Guchi-TP-015}
\date{\today
}
\title{
Quantum Scattering in Two Black Hole Moduli Space
}
\author{Kenji~Sakamoto${}^{1}$%
\thanks{e-mail: {\tt
b1795@sty.cc.yamaguchi-u.ac.jp}
} and
Kiyoshi~Shiraishi${}^{1,2}$%
\thanks{e-mail: {\tt
shiraish@sci.yamaguchi-u.ac.jp}
}}
\address{${}^1$Faculty of Science, Yamaguchi University\\
Yoshida, Yamaguchi-shi, Yamaguchi 753-8512, Japan
}
\address{${}^2$Graduate School of Science and Engineering,
Yamaguchi University\\
Yoshida, Yamaguchi-shi, Yamaguchi 753-8512, Japan
}
\maketitle
\begin{abstract}
We discuss the quantum scattering process 
in the moduli space consisting of two maximally 
charged dilaton black holes. 
The black hole moduli space geometry has different structures for 
arbitrary dimensions and various values of dilaton coupling. 
We study the quantum effects of 
the different moduli space geometries with 
scattering process. Then, it is found that there is a resonance state 
on certain moduli spaces. 
\end{abstract}
\section{Introduction}

Recently the study of black hole moduli space has attracted much attention. 
Quantum black holes have been studied by means of quantum fields 
or strings interacting with a single black hole. 
In the past several years the quantum mechanics of an arbitrary number $n$ of 
supersymmetric black holes has been focused upon. 
Configurations of $n$ static black holes parametrize a moduli space.
The low-lying quantum states of the system are governed by quantum mechanics 
in the moduli space. 
The effective theories in moduli space of 
Reissner-Nordstr\"{o}m multi-black holes were constructed 
in \cite{Gibbons}. 
Recently, (super)conformal quantum mechanics are 
constructed in the moduli space of 
four- and five-dimensional 
multi-black holes in \cite{jmas}, \cite{as2}. 
Further, we found that the quantum mechanics in moduli space of multi-black holes with dilaton coupling in diverse dimensions 
have the conformal symmetry \cite{Sakamoto:2002ci}. 

The scattering in moduli space was discussed by Ferrell and Eardley \cite{fe}, where the black hole-black hole scattering and coalescence in the moduli space, which is constructed from (3+1) dimensional Reissner-Nordst\"{o}m black holes, was calculated. 
The black hole moduli spaces geometry with dilaton coupling 
in $(N+1)$ dimensions was discussed by one of the present authors \cite{shir}. 
In these works the structure of the moduli space geometry is found to be 
different for dimensions and values of dilaton coupling. 

In this paper, we consider the quantum effects from the various moduli structures with the scattering in moduli space. 
In section 2, we discuss the moduli space metric of two black hole system 
with dilaton coupling in $(3+1)$ dimensions. 
In section 3, we investigate the quantum mechanics in the moduli space. 
From the feature of potential, we discuss whether an incoming particle is scattered away, or coalesces. 
In section 4, 
we consider the scattered-away processes. 
Using the WKB approximation, we calculate the phase shift and the deflection angle. 
Then we find the quantum effect in the moduli space. 
In the special cases of the moduli space, there are resonance states, which are discussed in section 5. 
In section 6, we give discussion and conclusion. 

\section{The Moduli Space Metric for the System of Maximally Charged 
Dilaton Black Holes}
In this section, we consider the moduli space of the Einstein-Maxwell-dilaton system along the technique discussed by Ferrell and Eardley \cite{fe}. 
The Einstein-Maxwell-dilaton system contains a dilaton field $\phi$ coupled to a $U(1)$ gauge field $A_{\mu}$ besides the Einstein-Hilbert gravity.
In the $(3+1)$ dimensions, the action for the fields with particle sources is 
\begin{equation}
S=\int d^{4}x \frac{\sqrt{-g}}{16\pi} \left[ R -
2 (\nabla \phi)^{2} - e^{-2a\phi} F^{2} \right]-\sum^n_{A=1} \int ds_A \left( m_A e^{-2a\phi} +Q_A {\bf A}\frac{d{\bf x}_A}{ds_A}\right),\label{eq:action}
\end{equation}
where $R$ is the scalar curvature and $F_{\mu\nu}=\partial_\mu A_\nu-\partial_\nu A_\mu$. $ds_A$ is the line element of the center of the $A$-th black hole. $m_A$ and $Q_A$ are the mass and the electric charge of the $A$-th black hole. We set the Newton constant $G=1$. 
The dilaton coupling constant $a$ can be assumed to be a positive value.

The metric for the $n$-body system of maximally-charged dilaton black holes has been known as \cite{shir}
\footnote{
In the special case of $a=0$ for $N=3$, the solution becomes the Papapetrou-Majumdar solution \cite{Papapetrou:1947ib}, \cite{Majumdar:eu}. Generally, these solutions have singularities. 
We, however, consider a kind of soliton 
with three long-range forces balanced (the gravitational force, the electric force and the dilatonic force) as the `black hole'. 
}
\begin{equation}
ds^2=-U^{-2}({\bf{x}}) dt^2+U^{2}({\bf{x}})d{\bf{x}}^2, 
\end{equation}
where
\begin{eqnarray}
U({\bf{x}})&=&(F({\bf{x}}))^{1/(1+a^2)}, \label{eq:U} \\
F({\bf{x}})&=&1+\sum^n_{A=1}\frac{\mu_A}{|{\bf{x}}-{\bf{x}}_A|}. \label{eq:F}
\end{eqnarray}

Using these expressions, the vector one form and dilaton configuration are written as
\begin{eqnarray}
A=\sqrt{\frac{1}{1+a^2}}\left(1-\frac{1}{F({\bf{x}})}\right)dt, \\
e^{-2a\phi}=(F({\bf{x}}))^{2a^2/(1+a^2)}.
\end{eqnarray}
In this solution, the asymptotic value of $\phi$ is fixed to be zero.

The electric charge $Q_A$ of each black hole are associated with the corresponding mass $m_A$ by
\begin{eqnarray}
m_A&=&\frac{1}{1+a^2}\mu_A,\\
|Q_A|&=&\sqrt{\frac{1}{1+a^2}}\mu_A. 
\end{eqnarray}
We consider that the perturbed metric and potential can be written in the form 
\begin{eqnarray}
ds^2=-U^{-2}({\bf{x}})dt^2+2{\bf{N}}d{\bf{x}}dt+U({\bf{x}})d{\bf{x}}^2, \\
A=\sqrt{\frac{1}{1+a^2}}(1-\frac{1}{F({\bf{x}})})dt+{\bf{A}}d{\bf{x}}, 
\end{eqnarray}
where $U({\bf{x}})$ and $F({\bf{x}})$ are defined by (\ref{eq:U}) and (\ref{eq:F}). 
We have only to solve linearized equations with perturbed sources up to $O(v)$ 
for $N_i$ and $A_i$. 
(Here $v$ represents the velocity of the black hole as a point source.) 
We should note that each source plays the role of a maximally charged dilaton black hole.

Solving the Einstein-Maxwell equations and substituting the solutions, 
the perturbed dilaton field and sources to the action (\ref{eq:action}) with proper boundary terms, 
we get the effective action up to $O(v^2)$ for $n$-maximally charged dilaton black hole system
\begin{eqnarray}
S&=&\int dt\left(\frac{1}{2}\sum^n_A m_A {\bf v}_A^2 \right) +\frac{3-a^2}{8\pi(1+a^2)^2}\int d^{4}x (F({\bf{x}}))^\frac{2(1-a^2)}{1+a^2} \nonumber \\
& &\hspace{.5cm} 
 \times \sum^n_{A,B=1}\frac{\mu_A\mu_B}{|{\bf{r}}_A|^{3}|{\bf{r}}_B|^{3}}\biggl[\frac{1}{2}({\bf{r}}_A \cdot {\bf{r}}_B)|{\bf{v}}_A-{\bf{v}}_B|^2-({\bf{r}}_A \times {\bf{r}}_B)\cdot({\bf{v}}_A \times {\bf{v}}_B) \biggl] , \label{lag1}
\end{eqnarray}
where ${\bf{r}}_A={\bf{x}}-{\bf{x}}_A$. $F({\bf{x}})$ is defined by (\ref{eq:F}).

In general, a naive integration in equation (\ref{lag1}) diverges. 
Therefore, we regularize the divergent terms proportional 
to $\int d^3x\delta^3(x)/|x|^p \ (p>0)$ 
which appear when the integrand is expanded must be regularized \cite{InPl}. We set them to zero. 

After regularization, the effective action for two body system (consisting of black holes labeled with $a$ and $b$) can be rewritten as
\begin{eqnarray}
S_{2B}=\int dt
\frac{1}{2} \mu {\bf{v}}^2 
\Biggl[ 1&-&\frac{M}{\mu}
-\frac{(3-a^2)M}{r}
+\frac{M}{m_a} \left( 1+(1+a^2)\frac{m_a}{r} \right) ^{(3-a^2)/(1+a^2)} \nonumber \\
& & \hspace{2cm} +\frac{M}{m_b} \left( 1+(1+a^2)\frac{m_b}{r} \right) ^{(3-a^2)/(1+a^2)}
\Biggl], 
\end{eqnarray}
where $M=m_a+m_b$, $\mu=m_am_b/M$, ${\bf{v}}={\bf{v}}_a-{\bf{v}}_b$ and 
$r=|{\bf{x}}_a-{\bf{x}}_b|$.
Thus the metric of the three dimensional moduli space for two-body system is 
\begin{equation}
g_{ij}=\gamma(r)\delta_{ij},\label{mmet1}
\end{equation}
with 
\begin{equation}
\gamma(r)= 1-\frac{M}{\mu}
-\frac{(3-a^2)M}{r}+\frac{M}{m_a} \left( 1+
\frac{(1+a^2)m_a}{r} \right) ^{\frac{3-a^2}{1+a^2}}+\frac{M}{m_b} \left( 1+
\frac{(1+a^2)m_b}{r} \right) ^{\frac{3-a^2}{1+a^2}}. \label{modmet2}
\end{equation}
The moduli space metric (\ref{mmet1}) depends on the dilaton coupling. 
We notice that the behavior of the moduli metric changes at $a^2=1/3$, 
where the moduli space has a critical structure. 
Therefore, the moduli space geometry has the various structure \cite{shir}. 

\section{Quantum mechanics in two-black hole moduli space} 
We consider the quantum scattering in this moduli space. 
The effective theories of quantum mechanics in moduli space of 
Reissner-Nordst\"{o}m black holes were constructed in \cite{TrFe}. 

Let us introduce a wave function $\Psi$ in the moduli space, which obeys the 
Schr\"{o}dinger equation
\begin{equation}
i\hbar\frac{d\Psi}{dt}=-\frac{\hbar^2}{2\mu}\nabla^2 \Psi, 
\end{equation}
where $\nabla^2$ is the covariant Laplacian constructed from the moduli space metric (\ref{mmet1}). 

The partial wave in a stationary state is 
\begin{equation}
\Psi=\psi_{ql}(r)Y_{lm}(\theta,\phi)\exp(-iEt/\hbar), \label{4scheq1}
\end{equation}
where $Y_{lm}(\theta,\phi)$ is the spherical harmonic function. 
We define the energy and variables as 
\begin{eqnarray}
E&=&\frac{\hbar^2 q^2}{2\mu}, \\
R&=&\int \sqrt{\gamma} dr, \\
\psi&=&\frac{\chi}{r\sqrt{\gamma}}.
\end{eqnarray}
The Schr\"{o}dinger equation (\ref{4scheq1}) is rewritten as 
\begin{equation}    
\frac{d^2\chi}{dR^2}+(q^2-V)\chi=0, \label{4scheq2}
\end{equation}
where the potential $V$ is represented as 
\begin{equation}
V=\frac{\gamma(r\gamma')'-r(\gamma')^2}{2r\gamma^3}+\frac{l(l+1)}{r^2\gamma}. \label{VVpot}
\end{equation}
Here $'$ stands for $\frac{d}{dr}$ and $l$ is the angular momentum. 

\begin{figure}[p]
\centering
\vspace*{2cm}
{\hspace{-.5cm}\includegraphics[width=6cm]{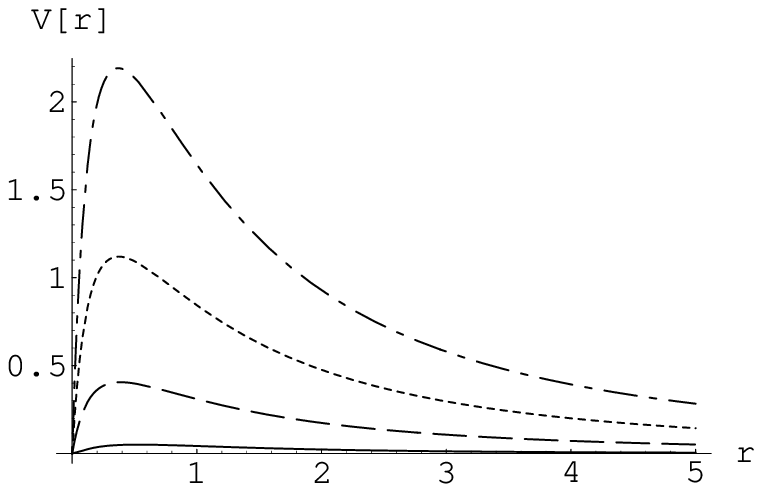}
\hspace{0.05cm}\includegraphics[width=6cm]{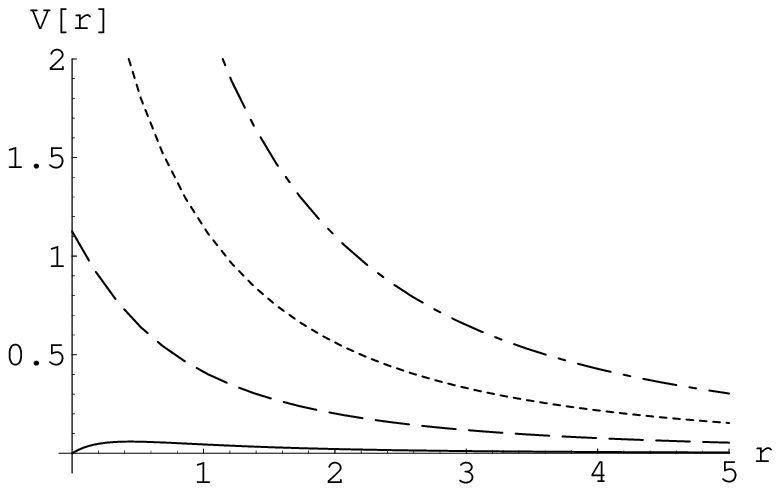}}\\ 
\vspace{-4.8cm}\hspace{-1cm}-A-\hspace{6.5cm}-B-\vspace{5.3cm} \\
\vspace{.5cm}
{\hspace{-.5cm}\includegraphics[width=6cm]{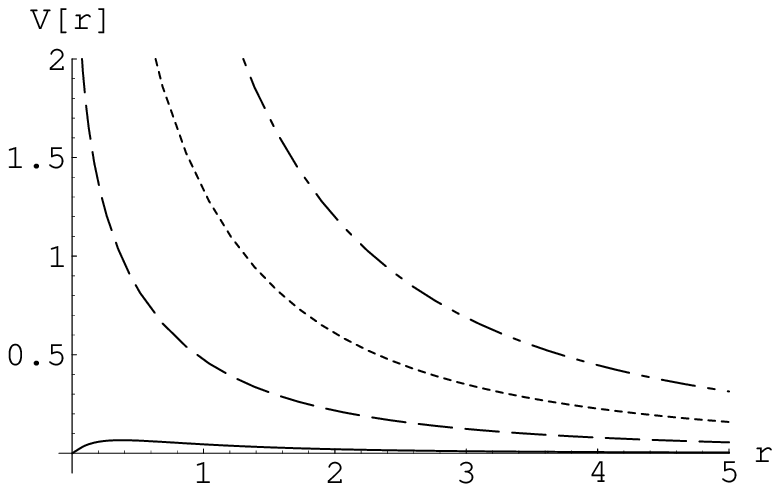}
\hspace{0.05cm}\includegraphics[width=6cm]{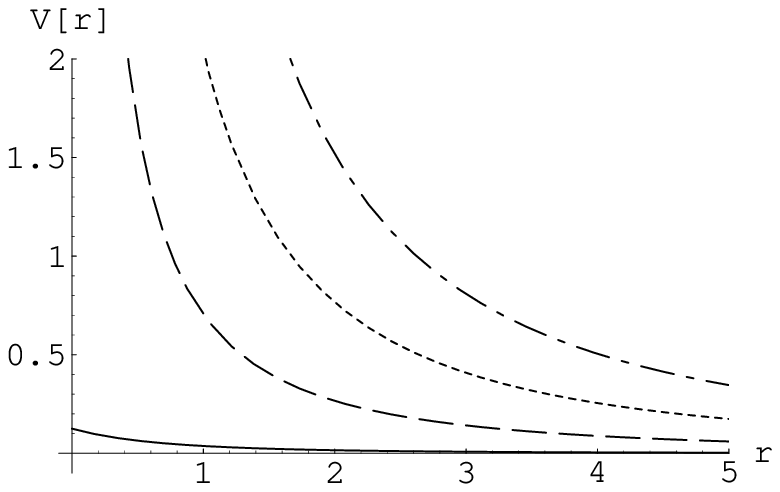}}\\ 
\vspace{-4.8cm}\hspace{-1cm}-C-\hspace{6.5cm}-D-\vspace{5.3cm} \\
\vspace{-2.cm}\includegraphics[width=5.5cm]{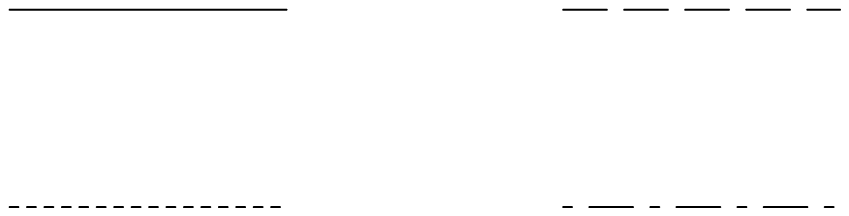} \\
\vspace{-2cm}\hspace{2.95cm} $l=0$ \hspace{2.2cm} $l=1$ \\ 
\vspace{.4cm}\hspace{2.95cm} $l=2$ \hspace{2.2cm} $l=3$ \\
\vspace{1cm}
\caption{The potential $V$ as the function of $r$ and angular 
momentum $l=0,1,2,3$, A for $a^2=0$, B for $a^2=1/3$, C for $a^2=1/2$ and D 
for $a^2=1$. We can find the potential barrier in the case of $a^2=0$ for all values of $l$ and $a^2=1/3, 1/2$ for $l=0$. }
\label{ff1}
\end{figure}

\begin{figure}[htpb]
\centering
\vspace*{2cm}
{\hspace{-.5cm}\includegraphics[width=7cm]{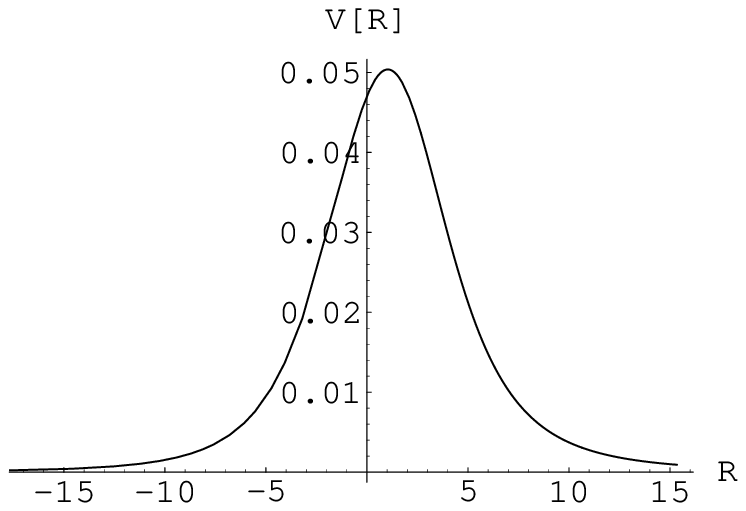}
\hspace{0.05cm}\includegraphics[width=7cm]{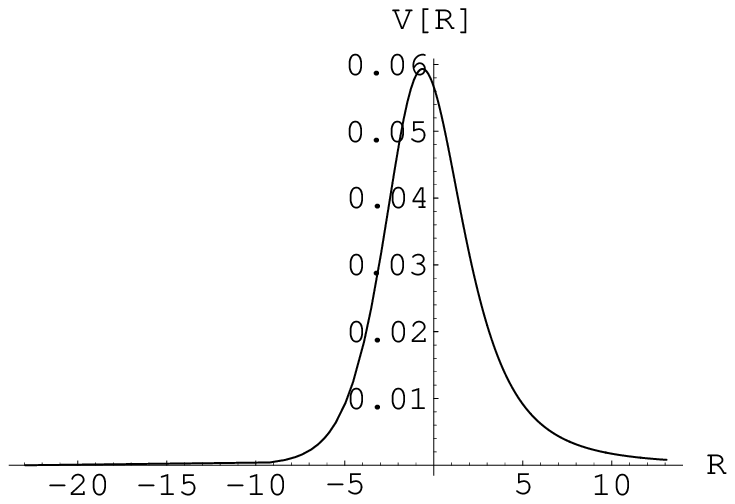}}\\ 
\vspace{-5.3cm}\hspace{-1cm}-A-\hspace{6.5cm}-B-\vspace{5.3cm} \\
\vspace{2cm}
{\hspace{-.5cm}\includegraphics[width=7cm]{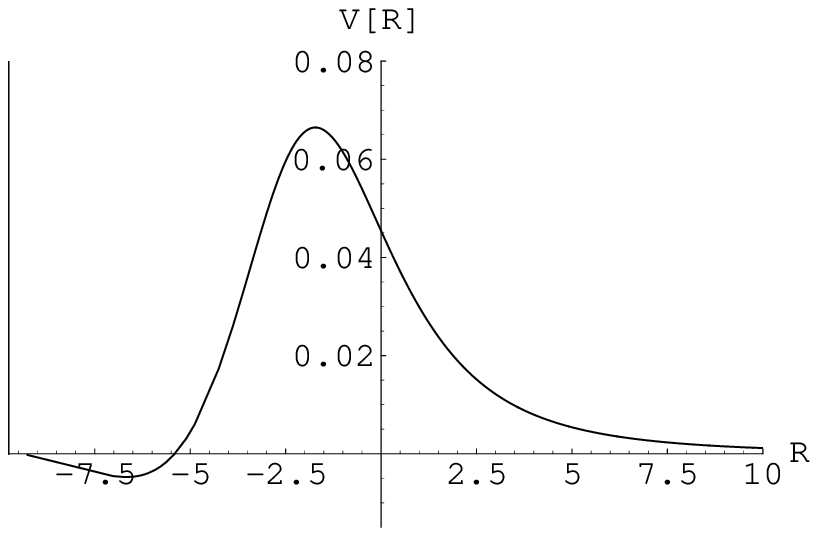}
\hspace{0.05cm}\includegraphics[width=7cm]{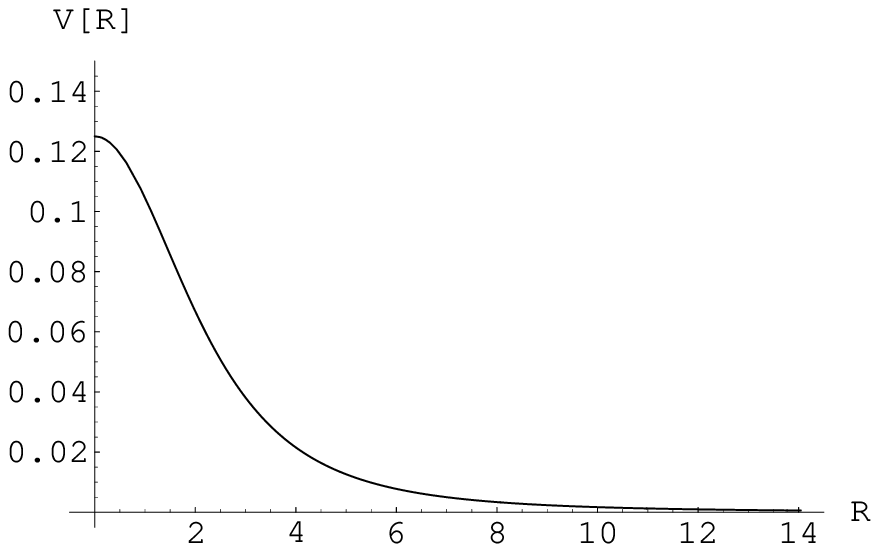}}\\ 
\vspace{-5.3cm}\hspace{-1cm}-C-\hspace{6.5cm}-D-\vspace{5.3cm} \\
\vspace{1cm}
\caption{The potential $V$ as the function of $R$ when the angular 
momentum $l=0$, A for $a^2=0$, B for $a^2=1/3$, C for $a^2=1/2$ and D for $a^2=1$.}
\label{ff13}
\end{figure}

To study the scattering process, we find the general view of the potential in the moduli space. The potentials for dilaton couplings $a^2=0$, $1/3$, $1/2$ and $1$ are plotted in Fig.~\ref{ff1} and Fig.~\ref{ff13}. 

From Fig.~\ref{ff1}, we can find that 
there is the potential barrier in the case of $a^2=0$ for all values of $l$ and $a^2=1/3, 1/2$ for $l=0$. 
If the potential barrier exists, the incoming particle approaching to the origin $r=0$ of the moduli space is scattered or 
coalesces. 
On the other hand, in the case of $a^2=1/3, 1/2$ for all values of $l\neq0$ 
and $a^2=1$ 
with any $l$, the incoming particle is always scattered away. 

Generally, if the angular momentum is zero $(l=0)$ the potential has the barrier in the case of $a^2<1$. For $a^2<1/3$, the variable $R$ diverges to minus infinity at the limit of $r\to0$. 
Then the particle going over the potential barrier 
can reach the minus infinity, $R\to-\infty$, so that, 
the incoming particle coalesces with the target black hole. 
In the case of $a^2>1/3$, the variable $R$ has the finite value at the limit of $r\to0$. 
The coordinate $R$ taking the infinite value means that 
the moduli space structure has the throat of infinite length \cite{shir}.
\footnote{
Exactly, the radial coordinate describing the moduli space structure 
in \cite{shir} is different from $R$. 
However, these behaviors at $r\to 0$ are similar. 
} 
In the case of $1/3<a^2<1$, 
the incoming particle through the potential barrier can reach the point of $r=0$ for the finite time and cannot stay at $r=0$. 
Therefore, the incoming particle is reflected at $r=0$ and scattered away out of the barrier, so that, does not coalesce. 
For the scattering of the incoming particle, we can guess 
that the potential becomes infinity at the point of $r=0$. 
The example of such a potential is shown in Fig.~\ref{ff13} C. 
We notice that the value of $a^2=1/3$ is the critical point of moduli structures \cite{shir}. 

\section{Scattering in two-black hole moduli space}
First, for simplicity, we will study the scattering-away processes in the case of $a^2=1/3$ and $a^2=1$ for arbitrary values of angular momenta. 
As the first study of quantum effects, we discuss 
the scattering process in the WKB approximation. 
In the WKB approximation, the phase shift in the scattering process is given by 
\begin{equation}
\delta_l=-qR_0+\int^{\infty}_{R_0}dR\left(\sqrt{q^2-V(R)-\frac{1}{4R^2}}-q\right)
+\frac{2l+1}{4}\pi,\label{delta}
\end{equation}
where $R_0$ is the solution of $V(R_0)=q^2-\frac{1}{4R_0^2}$. 
The partial cross section is 
\begin{equation}
\sigma_l=\frac{4\pi}{q^2}(2l+1)\sin^2\delta_l.
\end{equation}
The deflection angle is 
\begin{equation}
\Theta=\pi+\int^{\infty}_{R_0}dR\frac{\partial}{\partial l}\sqrt{q^2-V(R)-\frac{1}{4R^2}}.\label{Theta}
\end{equation}
These equations (\ref{delta})-(\ref{Theta}) are evaluated with respect to the outside of the potential barrier, where we ignore the effect of the inside of potential barrier. 
Thus, these are meaningful in the case of large $q$ when the potential barrier does not influence the calculation strongly. 
The low velocity and large $q$ can be realized if $\hbar \ll 1$, because $v \sim \hbar q/ \mu$. 

\begin{figure}[hbtp]
\centering
{\hspace{-1.5cm}\includegraphics[width=7.5cm]{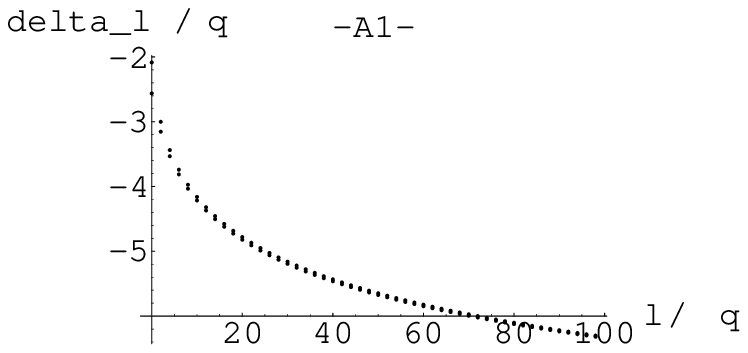}
\hspace{0.05cm}\includegraphics[width=7.5cm]{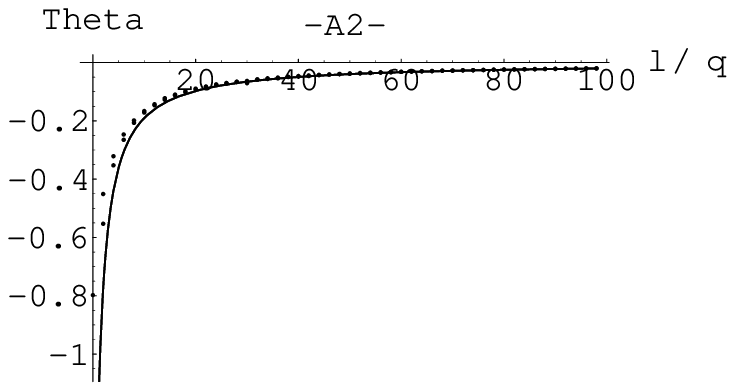}}\\ 
{\hspace{-1.5cm}\includegraphics[width=7.5cm]{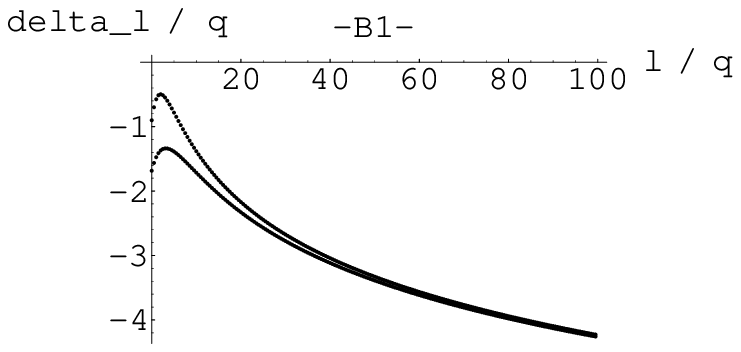}
\hspace{0.05cm}\includegraphics[width=7.5cm]{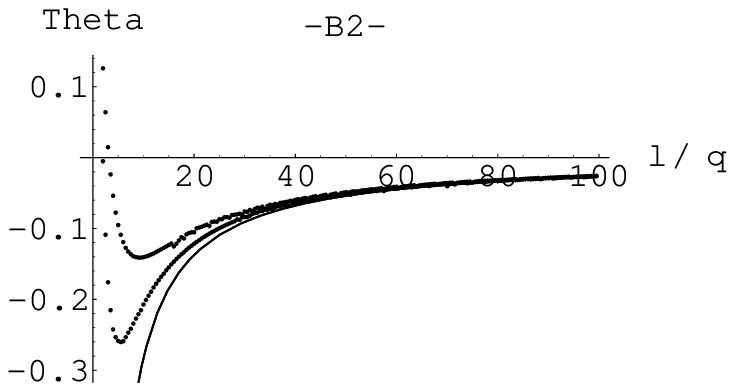}}\\ 
\caption{The phase shift and the deflection angle. A1 is a phase shift 
for $a^2=1$, A2 is a deflection angle for $a^2=1$, B1 and
 B2 are phase shift and deflection angle for $a^2=1/3$.}
\label{ff2}
\end{figure}

The phase shifts $\delta_l$ and the deflection angles
 $\Theta$ for $q=0.1$ and $0.2$ are plotted in Fig.~\ref{ff2}. 
The solid line in Fig.~\ref{ff2}~A2,~B2 represents the classical deflection 
angle which is obtained from the impact parameter \cite{KS3}. 
For $a^2=1$, the effects for the different value of $q$ are not found, and 
the deflection angle almost 
corresponds to the classical deflection angle. 
Although, for $a^2=1/3$, we can find that 
the behaviors of phase shift and deflection angle depend on 
the incoming particle energy. 
Then the deflection angle is 
different from the classical deflection angle at the small value of $l/q$. 
These effects are interpreted as the quantum effects 
arisen from the difference of the moduli space structures 
in the scattering process. 
Although we used the WKB semi-classical approximation in this analysis, 
the quantum effects can be obtained sufficiently. 

\section{Resonance}
\begin{figure}[hbtp]
\centering
{\includegraphics[width=8cm]{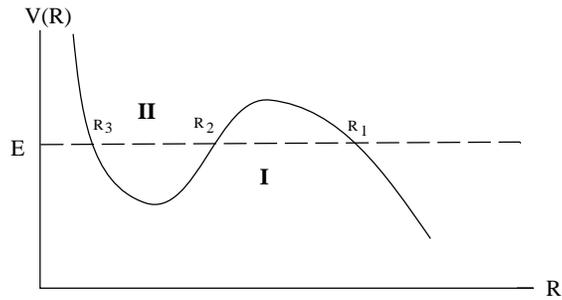}}
\caption{The form of typical Potential $V(R)$ with three turning points at 
$R=R_1, R_2, R_3$}
\label{zu}
\end{figure}

As seen in section III, the moduli space for $a^2>1/3$ has the finite region toward $r=0$, because the $R$ has a finite value in the limit $r\to0$ under the integration. 
The incoming particle through the potential barrier cannot stay at the point of $r=0$. 
Therefore, the coming particle is reflected at the point of $r=0$ and scattered away. 
We can guess that the potential becomes infinity at the point of $r=0$. 
Further, the potential for $a^2<1$ has the finite barrier. 
The example of the potential is shown in Fig.~\ref{ff13} C. 
The typical potential form for $1/3<a^2<1$ is shown in Fig.~\ref{zu}. 
The energy of an incoming particle is presented  by $E$ in Fig.~{\ref{zu}}.
At the point of $R=R_3$, the original variable $r$ equal to zero. 
For $1/3<a^2<1$, the potential has the barrier and three turning points. 
In this scattering process on the moduli space for $1/3<a^2<1$, 
we can expect that there is the resonance state.

In this section, we discuss the resonance state in the scattering process. 
The potential has three classical turning points. 
In Fig.~{\ref{zu}}, 
there is a classically allowed region II between $R_3$ and $R_2$ 
and for $R>R_1$ and 
a classically inaccessible region I between $R_2$ and $R_1$. 
In the WKB approximation, the scattering amplitude is 
given \cite{BrTaki}
\begin{equation}
S_l=\exp(2i \delta_1)\frac{1+\bar{N}(i\beta)\exp(2iW_{32})}{N(i\beta)+\exp(2iW_{32})},
\end{equation}
where $\delta_1$ is the WKB phase shift (\ref{delta}) evaluated with respect to the outermost turning point $R_1$, and 

\begin{eqnarray}
N(i\beta)&=&\frac{\sqrt{2\pi}\exp(-\pi\beta/2+i(\beta\ln\beta-\beta))}{\Gamma(\frac{1}{2}+i \beta)} ,\\
\bar{N}&=&N^* ,
\end{eqnarray}
with 
\begin{eqnarray}
\beta&=&\frac{i}{\pi} \int^{R_2}_{R_1} \sqrt{\chi(R)} dR
=\frac{1}{\pi} \int^{R_2}_{R_1}dR\sqrt{-(q^2-V(R)-\frac{1}{4R^2})}, \\ 
& &\hspace{2cm} W_{32}=\int_{R_3}^{R_2}dR\sqrt{q^2-V(R)-\frac{1}{4R^2}}.
\end{eqnarray}

\begin{figure}[hbtp]
\centering
\vspace*{1cm}
{\includegraphics[width=6cm]{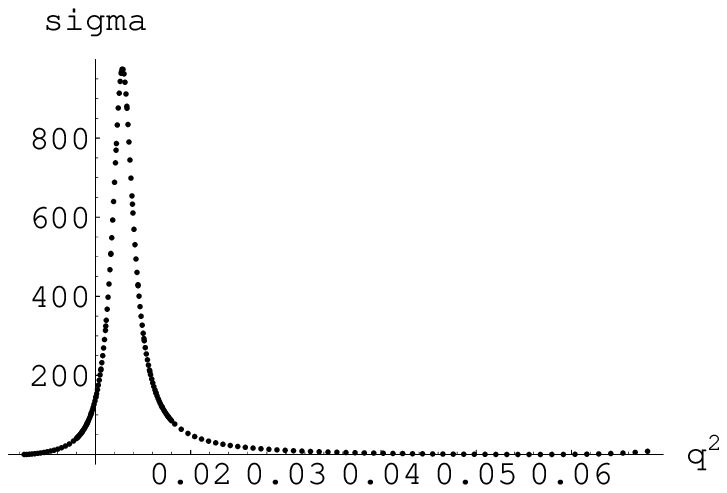}\hspace{.5cm} \includegraphics[width=6cm]{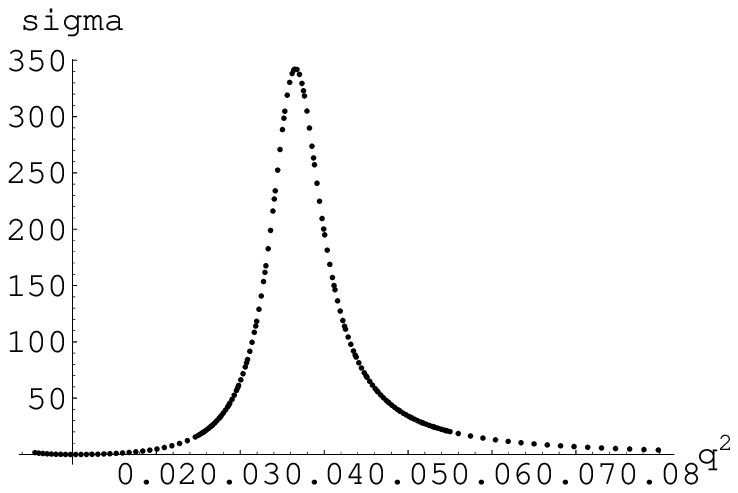}} \\
\vspace{-4.5cm}\hspace{0cm}{\small-A-}\hspace{6cm}{\small-B-}\vspace{4.5cm} \\
\vspace{.5cm}
{\small--C-}\\
\includegraphics[width=6cm]{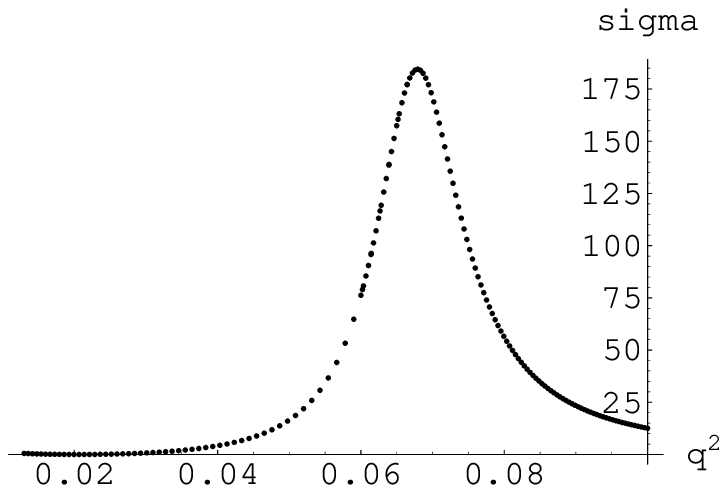} \\
\caption{The partial cross-section for $l=0$, 
A for $a^2=2/5$, B  for $a^2=9/20$ C for $a^2=1/2$.}
\label{cross}
\end{figure}

The partial cross section is represented by
\begin{equation}
\sigma_{l}=4 \pi(2l+1)|f_l|^2,
\end{equation}
where
\begin{equation}
f_l=\frac{1}{2iq}(S_l-1).
\end{equation}
In the cases of $a^2=2/5$, $a^2=9/20$ and $a^2=1/2$, 
the partial cross sections $\sigma_0$ are shown in Fig.~\ref{cross}.

Clearly, there is the resonance state. The quasi-bound state exists inside the potential barrier: The quasi-bound states for $a^2=2/5$ and $a^2=9/20$ 
exist below the maximal value of the potential barrier, $V\approx0.0619$ and $V\approx0.0641$ respectively. 
The quasi-bound state for $a^2=1/2$ exists near the maximal value
 $V\approx0.0665$. 

\begin{figure}[hbtp]
\centering
{\includegraphics[width=6cm]{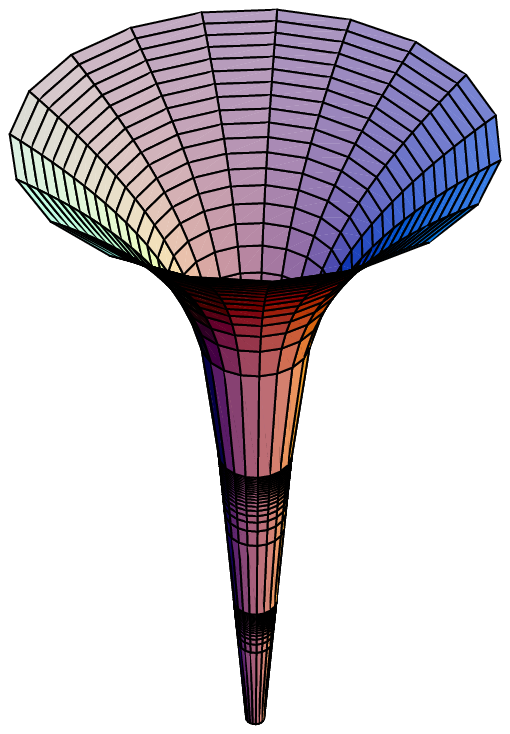}\includegraphics[width=5cm]{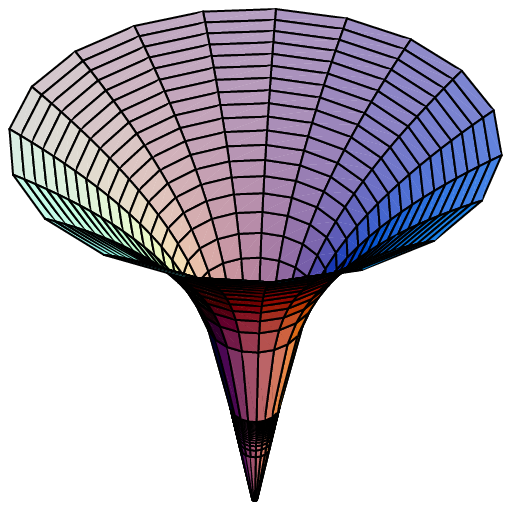}} \\
\vspace{-1.5cm}\hspace{-.5cm}-A-\hspace{5cm}-B-\\
\caption{The moduli space structure, 
A for $a^2=2/5$ and B for $a^2=1/2$.}
\label{modu4}
\end{figure}

Also, the width of resonance for $a^2=1/2$ is larger than for $a^2=9/20$ and $a^2=2/5$. 
Here the width of the resonance $\Gamma_0$ is given by
\begin{equation}
\Gamma_0=\frac{P}{T},
\end{equation}
where $P$ is the barrier penetration factor and $T$ is the period of the classical motion in the potential pocket inside the barrier: 
\begin{eqnarray}
P=(1+\exp{2\pi \beta})^{-1}, \\
T=\left(\frac{\partial W_{32}}{\partial E}\right)_{E_r},
\end{eqnarray}
where $E_r$ is the resonance energy. 
The width of the resonance is related to the time of classical motion of the test particle. 
From the width of resonance, 
we can guess that the time of the classical motion in moduli space for $a^2=2/5$ is longer than the case of $a^2=1/2$. 
Fig.~\ref{modu4} shows the surface of moduli space for $a^2=2/5$ and $1/2$ in the (3+1) dimensions. 
The moduli space geometry becomes deep and sharp for small values of $a^2$. 
Then, we suppose that the time of classical motion in moduli space for small $a^2$ is also long, because the trajectories of the classical motion becomes long. 

\section{Conclusion}
In this paper, we considered the quantum effects from the different moduli space in the scattering process. 
In three spatial dimensions, 
we obtained the phase shift and the deflection angle in the WKB approximation. 
Then we revealed the quantum effects in scattering process for $a^2=1/3$. 
We consider that these effects are due to the difference 
of the moduli space structure. 
Thus, the quantum effects from the moduli space structure are obtained in the scattering process. 
Moreover, there are the resonance states in the cases of $a^2=2/5, 9/20$ and $1/2$. 
These are the remarkable quantum effects, because there is no resonance state 
in the scattering process on the ordinary black holes with no dilaton coupling. 
The incoming particle falling into the ordinary black holes is simply absorbed or 
coalesce. 
In the scattering process in the moduli space for the specific case, however, there is the resonance state. 

In further study, we will consider the behavior in the different dimensions and for the other values of dilaton coupling $a$. 
The moduli space structures are different in the other dimensions and at the other values of dilaton coupling $a$. 
In (4+1) dimensions for $1<a^2<4$, the moduli space structure resembles 
the structure in (3+1) dimensions for $1/3<a^2<1$ \cite{shir}. Then we expect 
that the similar quantum effects are also arisen. 
The resonance is the particular state in the black hole moduli space. 
It seems that the effects of this resonance states on the quantum mechanics and conformal symmetry discussed in \cite{Sakamoto:2002ci} should be investigated. 
In this work, we considered the quantum effects in scattering process with the WKB approximation. 
We will discuss more precise quantum mechanical 
analysis in further study.

\newpage
\newpage
\vspace{1cm}
\newpage
\vspace{1cm}


\begin{thebibliography}{999}

\bibitem{Gibbons}
G.~W.~Gibbons and P.~J.~Ruback,
Phys.\ Rev.\ Lett.\  {\bf 57}, 1492 (1986).

\bibitem{jmas}
J.~Michelson and A.~Strominger,
JHEP {\bf 9909}, 005 (1999).

\bibitem{as2}
A.~Maloney, M.~Spradlin and A.~Strominger,
JHEP {\bf 0204}, 003 (2002).

\bibitem{Sakamoto:2002ci}
K.~Sakamoto and K.~Shiraishi,
Phys.\ Rev.\ D {\bf 66}, 024004 (2002).

\bibitem{fe} 
R.~C.~Ferrell and D.~M.~Eardley,
Phys.\ Rev.\ Lett.\  {\bf 59} 1617 (1987).

\bibitem{shir} 
K.~Shiraishi,
Nucl.\ Phys.\ B {\bf 402}, 399 (1993).

\bibitem{Papapetrou:1947ib}
A.~Papapetrou,
Proc.\ Roy.\ Irish Acad.\ (Sect.\ A)A {\bf 51}, 191 (1947).

\bibitem{Majumdar:eu}
S.~D.~Majumdar,
Phys.\ Rev.\  {\bf 72}, 390 (1947).

\bibitem{InPl} 
L.~Infeld and J.~Pleba\'nski, {\it Motion and Relativity}, 
(Pergamon Press, 1960).

\bibitem{TrFe} 
J.~Traschen and R.~Ferrell,
Phys.\ Rev.\ D {\bf 45}, 2628 (1992).

\bibitem{KS3}
K.~Shiraishi,
Int.\ J.\ Mod.\ Phys.\ D {\bf 2}, 59 (1993).

\bibitem{BrTaki} 
D. M. Brink and N. Takigawa,
 Nucl.\ Phys.\ A {\bf 279}, 159 (1977).

\end{thebibliography}
\end{document}